\documentstyle[aps,prd]{revtex} 
\tighten

\def\bfl{\begin{flushleft}}
\def\efl{\end{flushleft}}
\def\bfr{\begin{flushright}}
\def\efr{\end{flushright}}
\def\bc{\begin{center}}
\def\ec{\end{center}}
\def\be{\begin{equation}}
\def\ee{\end{equation}}
\def\ba{\begin{eqnarray}}
\def\ea{\end{eqnarray}}
\def\baa#1{\begin{array}{#1}}
\def\eaa{\end{array}}
\def\nn{\nonumber }
\def\lb#1{\label{#1}}

\def\drm{d}

\def\ap{{\tilde A}}
\def\up{{\tilde U}}
\def\mname{ELFIN }
\def\square{{\hbox{$\sqcup$}\llap{\hbox{$\sqcap$}}}}

\begin{document}

\wideabs{
\draft

\title{
Extended Lambda-Maxwell duality and related
large class of  dyonic and neutral
exactly solvable 
4D Einstein-Maxwell-dilaton models 
discovered 
      }

\author{
Konstantin G. Zloshchastiev
}

\address{Department of Physics, National University of Singapore,
Singapore 119260, Republic of Singapore\\
and 
Department of Theoretical Physics, Dnepropetrovsk State University,
Dnepropetrovsk 49050, Ukraine
}

\date{~Received: 12 Jan 2001 [arXiv]$^\dagger$ ~}
\maketitle

\begin{abstract}
We report the discovered class of exact static 
solutions of several 4D Einstein-Maxwell-dilaton systems:
string-induced, Liouville, trigonometric, polynomial, etc.,
for three basic topologies (spherical, hyperbolical and flat)
being uniformly treated.
In addition to the usual electric-magnetic duality
this class obeys a certain extended duality between 
Maxwell-dilaton coupling and dilaton mass.
Though major solutions we obtain are dyonic, 
the class also comprises interesting neutral 
models.
As a by-product, we significantly succeded
in resolving of the two important problems,
one of which has been standing
more than a decade (system with the string-inspired exponential
Maxwell-dilaton coupling and non-vanishing dilaton mass)
and another one - gravity coupled to massive neutral scalar field: 
generalized Liouville, 
Sin(h), Cos(h) - is about fifty years old.
Finally, we demonstrate the full separability of the static EMD system
and publicize the simple procedure of how to 
generate new integrability classes.

$^\dagger$
\tiny
Changes in arXiv versions: 
v3 - fixed misprint in eq. (25); v4 - slightly generalized eq. (5), 
added suggested refs; v5,6 - updated ref. \cite{txt-otherd}  and 
added important paragraph emphasizing
full separability and explaining how new integrability classes 
can be easily generated.

\end{abstract}

}


\narrowtext

Definitely, nowadays the Einstein-Maxwell-dilaton system,
described by the action 
\be
2 k_D^2 S = \int \drm^D x \sqrt{- g} \, Z
\biggl[
      R + B \,(\partial \phi)^2 + 
      \Xi \,F^2 + \Lambda 
\biggr],
                                                                 \lb{e-acti} 
\ee
with 
$Z$, $B$, $\Xi$ and $\Lambda$ being functions of $\phi$,
$F_{\mu\nu}= \nabla_\mu {\cal A}_\nu - \nabla_\nu {\cal A}_\mu$,
is the most important field-theoretical model, e.g.,
it appears in low-energy limit of string theory.
In general, the arena of such systems is the microworld 
where averaged charges cannot be made negligible.
In neutral case $\Xi \equiv 0$ this system is primarily
used in cosmology \cite{cosmo}
beginning from the (Jordan-Thirry-)Brans-Dicke models
or when studying fundamental aspects of black hole physics.
Anyway, theorists are highly interested in exact solutions (\ref{e-acti})
because no satisfactory perturbation theory has been constructed 
for it and numerical solutions can be regarded only as additional arguments.
To settle all the conventions let us first extract field
equations from the action above:
\ba
&&G_{\mu\nu} -  g_{\mu\nu}
\biggl[
\left(
      \frac{B}{2} - \frac{Z_{,\phi\phi}}{Z}
\right) (\partial \phi)^2 +
\frac{\Xi\, F^2 + \Lambda}{2} -
\frac{Z_{,\phi}}{Z} \square \phi
\biggr]
\nn\\ &&\ \
+
\left(
      B- \frac{Z_{,\phi\phi}}{Z}
\right) \partial_\mu \phi \, \partial_\nu \phi +
2\, \Xi\, F_{\mu\alpha} F_{\nu}^{\ \alpha} =
\frac{Z_{,\phi}}{Z} \nabla_\mu \nabla_\nu \phi \ ,              \nn\\
&&2\,Z B\, \square \phi +
(Z B)_{,\phi} (\partial \phi)^2
- 
Z_{,\phi} R -
(Z \Xi)_{,\phi} F^2 =
(Z \Lambda)_{,\phi} \ ,                                          \nn\\
&& Z \Xi \, \nabla^\mu F_{\mu\nu} +
(Z \Xi)_{,\phi} 
 \partial^\mu \phi \, F_{\mu\nu} =0,  \nn
\ea
where $G$ is the Einstein tensor, 
the subscript ``$,\phi$'' stands for the derivative
with respect to scalar field.
Without loss of generality we can assume $Z(\phi)=1$ 
(that 
can be always achieved 
in $D>2$)
and $B(\phi)=-\beta/2$, $\beta$ is a constant -
we wish to keep it unfixed as a regulator of the dilaton's rescaling which
may involve the imaginary unit.
We are interested in 4D static solutions hence further we will work
with the metric ansatz
\ba
&&
\drm s^2 = - e^{U(r)} \drm t^2 + e^{-U(r)} \drm r^2 +
e^{A(r)} \drm \Omega^2_{(k)}, \\
&&
\drm \Omega^2_{(k)} \equiv 
\left\{
\baa{ll}
\drm \theta^2 + \text{sin}^2 \theta \, \drm \varphi^2, 
&k=1 , \\
\drm \theta^2 +  \theta^2 \drm \varphi^2, 
&k=0 , 
\\
\drm \theta^2 + \text{sinh}^2 \theta \, \drm \varphi^2, 
&k=-1 , 
\eaa
\right.
\ea
thus $k$ enumerates these three topologies - spherical, flat 
and hyperbolic  -
we will work with all of them simultaneously and uniformly.
Similarly, the electromagnetic potential 1-form
is assumed as
\be
{\cal A} = 
\left\{
\baa{ll}
 \omega (r)\, \drm t - P\, \text{cos}\,\theta \,\drm \varphi, 
&k=1 , \\
 \omega (r)\, \drm t - \frac{1}{2} P\,\theta^2 \drm \varphi, 
&k=0 , \\
 \omega (r)\, \drm t - P\, \text{cosh}\,\theta \,\drm \varphi, 
&k=-1 , 
\eaa
\right.
\ee
with constant $P$ being the  magnetic charge.
With all this in hand the field equations above
take the form 
(we adopt curvature conventions from Appendix of 
Ref. \cite{cllp}):
\ba
&&
A'' + A' (A' + U') - 
 \hat\Xi \, e^{-2 A-U} 
=
\Lambda e^{-U} + 2 k e^{-A-U}
,                                                        \lb{e-em1}\\
&&
U'' + U' (A' + U') +  
 \hat\Xi \, e^{-2 A-U} =
\Lambda e^{-U} 
,                                                         \lb{e-em2}\\
&&
\beta  \phi'' + \beta \phi' (A' + U') +  
\hat\Xi_{,\phi}\, e^{-2 A-U} +
\Lambda_{,\phi}\, e^{-U} 
=0,                                                       \lb{e-em3}  \\
&&
2 A'' + A'^2 + \beta \phi'^2
=0,                                                        \lb{e-em4} \\
&&
\Xi \,\omega' = Q e^{-A} ,                            \lb{e-em5}
\ea
where 
$' \equiv \partial_r$,
$\hat\Xi \equiv 2 \, (Q^2\,\Xi^{-1} + P^2 \, \Xi )$,
integration constant $Q$ stands for electric charge
up to a coefficient.

Now we are in position to present the promised class  of solutions
which was obtained using the 
approach \cite{zlo-way}.
Thus, if $\Xi (\phi)$ and $\Lambda (\phi)$ obey
the linear ODE which is
invariant under extended duality 
transformations
$
\{
\Lambda \leftrightarrow \hat\Xi, \,
d_2 \leftrightarrow 1/d_2, \,
d_1 \leftrightarrow -d_1, \,
\beta \leftrightarrow -\beta
\},
$
in addition to the usual electric-magnetic 
$
\{
\Xi \leftrightarrow 1/\Xi, \,
Q \leftrightarrow P
\} 
$
duality
ones, 
\ba
&&
  \frac{e^{d_1 \phi}}{d_2}\,
   \left( \Lambda  + 
     \frac{d_1}{\beta } \,\Lambda_{,\phi}
     \right)
+
\frac{d_2}{ e^{d_1 \phi}}
\left(
      \hat\Xi + \frac{d_1}{\beta}\, \hat\Xi_{,\phi}
\right)  
=- 2\,k,                                    \lb{e-cl1a}
\ea
where 
$d_1, \, d_2 \equiv e^{d_1 \phi_0}$ are arbitrary constants,
then 
\be                                                   \lb{e-cl1b}
e^A =
r^{\frac{2\,d_1^2}{\beta  + d_1^2}}/d_2 , \
e^\phi = r^{\frac{2 d_1}{\beta + d_1^2}},
\ee
as for $U$ and $\omega$ then their problem is trivial \cite{txt-uw}.
Here we will not address the separate good issue -
does the extended Lambda-Maxwell duality above play any special
physical role, especially in $D=10$ and $11$ \cite{txt-otherd}.
It may happen that such dualities can be not only
the abstract generators of  classes of integrability but
also can shed light upon the fundamental nature
of the cosmological (dilaton mass) term $\Lambda$ 
which is precisely known
neither in string theory nor in cosmology.

The class (\ref{e-cl1a}), (\ref{e-cl1b}) has a singular sub-branch if
$\beta + d_1^2 = 0$. If we choose the plus root  then 
Eq. (\ref{e-cl1a}) holds with $d_1  = i \sqrt \beta$
but instead of (\ref{e-cl1b}) we have
\be                                                   \lb{e-cl1sb}
A = i \chi r - \ln d_2, \
\phi = \chi r/\sqrt\beta , 
\ee
where $\chi$ is another arbitrary constant.
Thus, the expressions (\ref{e-cl1a}) and (\ref{e-cl1b}) or 
(\ref{e-cl1a}) and (\ref{e-cl1sb})
(with the recipe \cite{txt-uw} kept in mind) 
yield a complete general-in-class solution.
Now, to demonstrate how useful and large this class is, let us consider
its most key or important specimens.
At the same time we will resolve some problems that 
harassed theorists for decades.

~\\
({\it a}) {\it Gravity coupled to neutral scalar field}: $\Xi \equiv 0$.

Historically the neutral scalar field is the oldest system 
(massless neutral scalar field was considered by Fisher in 1948 \cite{fis48}
and rediscovered afterwards by JNW \cite{jnw68})
so let us see what is the mass term our class yields.
Integrating of Eq. (\ref{e-cl1a}) at $Q=P=0$ 
reveals the following three cases:

({\it a.1}) $\beta-d_1^2 \not= 0$.\\
We obtain the (generalized) Liouville model
\be                                                          \lb{e-a1gen}
\Lambda=
a_2 e^{-\frac{\beta\phi}{d_1}} -
\frac{2 k \beta d_2}{\beta-d_1^2} e^{-d_1 \phi},
\ee
where $a_2$ is another arbitrary constant,
and from  Eq. (\ref{e-cl1b}) and Ref. \cite{txt-uw} we have
the  two following subcases:

({\it a.1.1}) $\beta-3 d_1^2 \not= 0$.
Here $\Lambda$ is as above 
whereas the solution is given precisely
by Eq. (\ref{e-b11}) at $Q=P=0$.

({\it a.1.2}) $\beta-3 d_1^2 = 0$.
Choosing for definiteness
the root $d_1 = \sqrt{\beta/3}$ we have
\ba
\Lambda = 
e^{-\sqrt{3\beta}\, \phi} 
\left(
a_2 - 3 k d_2 \,
e^{2\sqrt{\beta/3}\, \phi} 
\right),
\ea
so this is special case $Q=P=0$ of the solution (\ref{e-b12}).

({\it a.2}) $\beta + d_1^2 = 0$.\\
The $\Lambda$ in this case is a special case of 
Eq. (\ref{e-a1gen}) but we want to treat it separately because
this is the only way to combine exponents into trigonometric
functions,
and this is an example where the usefulness
of the sub-class (\ref{e-cl1sb}) becomes evident.
Using Eq. (\ref{e-cl1sb}) and \cite{txt-uw} we obtain
\be
\Lambda = 
a_2 \,
e^{i\sqrt{\beta}\, \phi} 
 - \frac{k d_2}{
e^{i \sqrt{\beta}\, \phi}},   \
e^U = 
\frac{ k\, d_2 r + c}{i \chi\, e^{i \chi r}}
-\frac{a_2 e^{i \chi r}}{2 \chi^2}
, 
\ee
where 
$c$ is another arbitrary constant
and
$A$ and $\phi$ are given by Eq. (\ref{e-cl1sb}).
The complete story is that at $k \not= 0$ 
this solution contains the simplest
trigonometric potentials.
If we assume
$
a_2 = - k d_2, 
$
then we come to the Cosine-Einstein model
and
similarly for Sine-Einstein
($a_2 = k d_2$, $d_2 = i\, \bar d_2$):
\be
\Lambda = 
- 2 k d_2 \cos{(\sqrt{\beta}\, \phi)}, \ \
\Lambda = 
- 2 k\, \bar d_2 \sin{(\sqrt{\beta}\, \phi)},  
\ee
and for Sinh and Cosh just by Wick rotation.
The solutions of such models
on flat or fixed background geometry
(Sine-Gordon, etc.) have been got long ago
but to our knowledge so far nobody has managed to 
obtain any self-gravitating solution despite tremendous efforts
were
made in view of evident importance of the subject.

({\it a.3}) $\beta-d_1^2 = 0$.\\
Choosing for definiteness
a positive root we 
obtain the Liouville model coupled to a linear term
\be
\Lambda = 
e^{-\sqrt\beta \phi}
\left(
a_2 - 2 k \sqrt\beta \, d_2 \, \phi
\right)
,   
\ee
its solution is given by Eq. (\ref{e-b2}) at zero charges,
and with this model the neutral case is exhausted. 

~\\
({\it b}) {\it String-inspired coupling}: $\Xi = a_1 e^{b_1 \phi}$.

The physically interesting cases are as follows (but not limited
to).
If one assumes in action (\ref{e-acti}) 
that ($D=4$): $\beta=16/(D-2)$, $a_1 = - k_D^2/2$,
$b_1 = -4 g_2 / (D-2)$
then: $g_2 = 1$ corresponds to field theory limit of superstring model
(more precisely, compactified effective theory if $D=4$),
$g_2 = \sqrt{1+\frac{D-2}{n}}$ corresponds to the toroidal $T^n$ reduction
of ($D+n$)-spacetime to $D$-spacetime,
$g_2 = 0$ is a usual Einstein-Maxwell system.
The previously done work is: 
Gibbons and Maeda \cite{gm88} received 
the single-charged 
solutions for $\Lambda = 0$ and arbitrary $D$ and $g_2$,
Dobiasch and Maison \cite{dm82} discovered the dyonic solutions for
$g_2 = \sqrt 3$, see also Refs. \cite{gib82,ghs91,gw86},
several solutions for potentials of special type were
obtained in works \cite{cllp,sev-b,sev-exp}.
Also a lot of qualitative and numerical work 
has been done 
\cite{qua-num,hh93,gh93}.
Summarizing what has been achieved one must say that major solutions
were received when $\Lambda =0$ 
(except Refs. \cite{cllp,sev-exp}) 
and in single-charged cases whereas
$\Lambda \not=0$ and dyonic ones still remain a {\it terra incognita}.
Such a situation 
does not seem to be satisfactory because in string-induced
theories $\Lambda$ is not negligible as a rule (though unknown).
Let us turn to our class.
Integrating of Eq. (\ref{e-cl1a}) 
reveals the following cases.

({\it b.1}) $\beta-d_1^2 \not= 0$, 
$\beta \pm b_1 d_1 - 2 d_1^2 \not= 0$.\\
Then the dilaton mass term is given by
\ba
&&
\Lambda = 
a_2 e^{-\frac{\beta}{d_1}\phi}
- \frac{2 k \beta d_2}{\beta - d_1^2} e^{-d_1 \phi}
- \frac{2 a_1 d_2^2 P^2\, (\beta + b_1 d_1)}{\beta + b_1 d_1 - 2 d_1^2}
\nn\\&&\qquad
\times
e^{(b_1 - 2 d_1)\, \phi}
- \frac{2 d_2^2 Q^2\, (\beta - b_1 d_1)}{a_1 \,(\beta - b_1 d_1 - 2 d_1^2)}
e^{-(b_1 + 2 d_1)\, \phi}.              \lb{e-lamb1}
\ea
One should not be concerned 
with appearance of $Q$ and $P$ - it simply
means that these charges are not arbitrary but depend on constants
in front of corresponding exponents
(for the sake of uniformity we do not make additional 
redefinitions of newborn arbitrary constants 
$a_2, \, d_1, \, d_2$, etc.).
Incidentally, note that $\Lambda$ 
is essentially exponential hence it can be either
large or incredibly small value - 
the former takes place in microworld whereas the latter does in cosmology.
Further, when obtaining $U$ 
we reveal a number of additional subcases 
generalizing Ref. \cite{sev-exp}:

({\it b.1.1}) $\beta-3 d_1^2 \not= 0$, $\beta + d_1^2 \not= 0$,
$\beta \pm 2 b_1 d_1 - d_1^2 \not= 0$.
Then $\Lambda$ is given by Eq. (\ref{e-lamb1}) whereas the complete
solution is
\ba
&&
e^U = 
\frac{(\beta + d_1^2)^2}{2 d_1^2}
\Biggl[
c \, r^{1 - \frac{2\,d_1^2}{\beta  + d_1^2}}
- 
\frac{2\,k d_1^2\,d_2 \, r^{\frac{2\,\beta }{\beta  + d_1^2}}}
       {{\beta }^2 - d_1^4} 
\nn\\&&\quad
- 
\frac{a_2 \,r^{\frac{2\,d_1^2}{\beta  + d_1^2}}
       }
       {\beta  - 3\,d_1^2} 
- 
  \frac{4\,P^2\,
        a_1\, d_1^2 \, d_2^2 \,
        r^{\frac{2 \left( \beta  + b_1 d_1 -d_1^2 \right) }
                {\beta  + d_1^2}
          }
       }{( \beta  +  b_1 d_1 - 2\,d_1^2 ) 
         ( \beta  + 2\,b_1\,d_1 - d_1^2 ) } 
\nn\\&&\quad
- 
\frac{4\,Q^2\,d_1^2\,d_2^2 \,
      r^{\frac{2\,\left( \beta  - b_1 d_1 - d_1^2 \right) }
        {\beta  + d_1^2}}
     }
     {a_1
      ( \beta  - b_1\,d_1 - 2\,d_1^2 ) 
      ( \beta  - 2\,b_1\,d_1 - d_1^2 ) 
     }
\Biggr],
\nn\\&&
\omega - \omega_0 =
\frac{
        Q\,
        d_2
        \left( \beta  + d_1^2 \right) \,
       }
      {{a_1}\,\left( \beta  - 2\,b_1\,d_1 - d_1^2 \right) } 
        r^{1 - 
           \frac{2\,d_1 \left(b_1 + d_1 \right)}
              {\beta  + d_1^2}
          },                                       \lb{e-b11}
\ea
and $A$ and $\phi$ are given precisely by Eq. (\ref{e-cl1b}).

({\it b.1.2}) $\beta-3 d_1^2 = 0$, $\beta + d_1^2 \not= 0$,
$\beta \pm 2 b_1 d_1 - d_1^2 \not= 0$.  
Choose the positive root then 
$\Lambda$ is given by Eq. (\ref{e-lamb1}) at $d_1 = \sqrt{\beta/3}$
and the solution is
\ba
&&
e^U = 
2 \sqrt r
\Biggl[
c + a_2 \ln r 
- k\,{d_2} r 
-
\frac{8\,
      \beta \,d_2^2 \,Q^2
      r^{\frac{1}{2}
         \left(
               1 - \sqrt{\frac{3}{\beta}} b_1
         \right)  
        }
     }
     {{a_1}\,{\left( {\sqrt{\beta }} - {\sqrt{3}}\,b_1 \right) }^2} 
\nn\\&& \qquad
- 
\frac{8\,
        \beta \,{a_1}\,
        d_2^2 \,P^2
      r^{\frac{1}{2}
         \left(
               1 + \sqrt{\frac{3}{\beta}} b_1
         \right)  
        }
     }
     {{\left( {\sqrt{\beta }} + {\sqrt{3}}\,b_1 \right) }^2}
\Biggr],
\quad
e^A =
\frac{\sqrt r}
     {d_2} 
,
\nn\\&&
e^{2\phi} =
r^{\sqrt{\frac{3}{\beta}}} ,
\
\omega - \omega_0 =
\frac{2 \sqrt \beta\, d_2 \, Q / a_1}{\sqrt \beta - \sqrt 3 \, b_1}
      r^{\frac{1}{2}
         \left(
               1 - \sqrt{\frac{3}{\beta}} b_1
         \right)  
        }
.                                                       \lb{e-b12}
\ea
Other ({\it b.1})'s subcases when expressions (\ref{e-b11}) but
not (\ref{e-lamb1}) become singular can be treated similarly using
Ref. \cite{txt-uw} and either
Eq. (\ref{e-cl1b}) or (\ref{e-cl1sb}).
Their general feature is the appearance of logarithm in $e^U$.

({\it b.2}) $\beta-d_1^2 = 0$, 
$\beta \pm b_1 d_1 - 2 d_1^2 \not= 0$.    \\
We choose the root $d_1 = \sqrt\beta$
then following Eq. (\ref{e-cl1a})
the dilaton mass term is given by
\ba
&&
\Lambda = 
(a_2 - 2 k \sqrt\beta d_2\, \phi) e^{-\sqrt\beta \phi}
+ 2 d_2^2 e^{-2\sqrt\beta\, \phi} 
\nn\\&&\qquad
\times
\biggl[
       \frac{a_1 P^2 (\sqrt\beta + b_1)}{\sqrt\beta - b_1}
       e^{b_1 \phi}
       +
       \frac{Q^2 (\sqrt\beta - b_1)}{a_1(\sqrt\beta + b_1)}
       e^{-b_1 \phi}
\biggr],
\lb{e-lamb2}
\ea
and the only solution (no subcases) is
\ba
&&
e^U = 
c 
- 2 k\, d_2 \,r\, (\ln r -2)
+ 
\frac{4 \beta\, d_2^2}{b_1}
\biggl[
\frac{a_1 P^2}{\sqrt\beta - b_1}
r^{\frac{b_1}{\sqrt\beta}}
\nn\\&&\qquad
-
\frac{Q^2/a_1}{\sqrt\beta + b_1}
r^{-\frac{b_1}{\sqrt\beta}}
\biggr]
+ a_2 r 
,
\  \
e^A = r/d_2 
,
\nn\\&&
e^{\sqrt\beta\phi} = r,
\ \
\omega - \omega_0 =
\frac{\sqrt\beta \, d_2 \, Q}{a_1 \, b_1}
      r^{-\frac{b_1}{\sqrt\beta}}
.                                              \lb{e-b2}
\ea

({\it b.3}) $\beta-d_1^2 \not= 0$, 
$\beta + b_1 d_1 - 2 d_1^2 = 0$,
$\beta - b_1 d_1 - 2 d_1^2 \not= 0$.\\
To avoid wearisome square-root branches
let us impose 
$
b_1 = 2 d_1 - \beta/d_1,
$
and 
work with $d_1$ bearing in mind its relation to the given parameter
$b_1$.
We have
\ba
&&
\Lambda = 
\left(
a_2 -
4 a_1 d_1 d_2^2 P^2 \, \phi
\right)
e^{-\frac{\beta}{d_1} \phi}
- \frac{2 k \beta d_2}{\beta-d_1^2}
e^{- d_1 \phi}
\nn\\&&\qquad
- 
\frac{2 d_2^2 \, Q^2 (\beta-d_1^2)}{a_1 (\beta - 2 d_1^2)}
e^{\frac{\beta-4 d_1^2}{d_1} \phi}
,
\ea
whereas the solution (at $3 \beta - 5 d_1^2 \not=0$)
is expressed as
\ba  
&&e^U = 
(\beta+ d_1^2)
\Biggl[
f\, 
r^{\frac{2 d_1^2}{\beta + d_1^2}}
+
c \,
r^{1-\frac{2 d_1^2}{\beta + d_1^2}}
-
\frac{k d_2
      r^{\frac{2 \beta}{\beta + d_1^2}}
      }{\beta - d_1^2}
\nn\\&&\quad
-
\frac{d_2^2 (\beta + d_1^2) Q^2 
      r^{\frac{4 \beta - 6 d_1^2}{\beta + d_1^2}}
     }
     {a_1 (3\beta - 5 d_1^2) (\beta - 2 d_1^2)}
\Biggr]
,\ \ 
\omega  =
\frac{d_2 Q (\beta + d_1^2)}{a_1 (3\beta - 5 d_1^2)}
r^{\frac{3\beta - 5 d_1^2}{\beta + d_1^2}}
,
\nn
\ea
with $A$ and $\phi$ given by Eq. (\ref{e-cl1b}),
where we have denoted
\ba
&&
f \equiv
- \frac{\beta + d_1^2}{d_1^2 (\beta - 3 d_1^2)}
\biggl[
\frac{a_2}{2}
+
\frac{a_1 d_2^2 P^2
     }
     {(\beta + d_1^2) (\beta - 3 d_1^2)
     }
\nn\\&&\qquad \
\times
      \left(
            \beta^2 - 2 \beta d_1^2 (2 \ln r +3)
            + d_1^4 (12 \ln r - 7)
      \right)
\biggr].  \nn
\ea
The subcase $3 \beta - 5 d_1^2 =0$, i.e., 
$\{b_1,\, d_1\} = \{\pm \sqrt{\beta/15}, $ $ \pm\sqrt{3\beta/5}\}$,
can be treated similarly (nothing special with it
except that $\omega$ turns out to be a linear function of $\ln r$).

({\it b.4}) 
$\beta-d_1^2 \not= 0$, 
$\beta + b_1 d_1 - 2 d_1^2 \not= 0$,
$\beta - b_1 d_1 - 2 d_1^2 = 0$.\\
This case is in some sense a counterpart of the previous one
({\it b.3}) but the roles of $Q$ and $P$ are interchanged.
It can be treated absolutely similarly as well as
other cases when expressions
$\beta-d_1^2$, 
$\beta \pm b_1 d_1 - 2 d_1^2$,
are equal zero pairwise (the case when they are zeros all together
is inconsistent).
In all these cases $\Lambda$'s resemble that from ({\it b.3}), i.e., are the
combinations of exponents of dilaton coupled to 
linear functions of dilaton. 

({\it b.5}) $\beta+d_1^2 = 0$.\\
We choose the plus root then $\Lambda$ is given by Eq. (\ref{e-lamb1})
at $d_1=i\sqrt\beta$
whereas the solution 
is 
(provided $b_1 \not= \pm 3 i \sqrt\beta$)
\ba  
&&-\chi^2 e^{ U} =
\frac{a_2}{2} e^{i \chi r}
+ 
\frac{i \chi (c+ k d_2 r)}{e^{i \chi r}} 
-
\frac{
      4 \beta d_2^2  
      e^{-2 i \chi r}
     }{a_1}
\nn\\&&\ \
\times
\biggl[
\frac{a_1^2 P^2 e^{\frac{b_1 \chi}{\sqrt\beta} r}}
     {(\sqrt\beta + i b_1) (3 \sqrt\beta + i b_1)}
+
\frac{Q^2 e^{-\frac{b_1 \chi}{\sqrt\beta} r }}
     {(\sqrt\beta - i b_1) (3 \sqrt\beta - i b_1)}
\biggr]
,\ \ 
\nn\\&&
\omega - \omega_0 =
\frac{ i \sqrt\beta d_2 Q}{a_1 \chi (\sqrt\beta - i b_1)}
e^{-\chi r \bigl( i + \frac{b_1}{\sqrt\beta} \bigr)}
,
\ea
and $A$ and $\phi$ are given by Eq. (\ref{e-cl1sb}).
The subcases $b_1 = \pm i \sqrt\beta$,
$\pm 3 i \sqrt\beta$ can be done by analogy.
Instead of handling it let us outline some other interesting
$\Xi \, \Lambda$ pairs that belong to the class.
If thereby we do not list $U$'s and $\omega$'s then 
a reader is referred to 
Ref. \cite{txt-uw}.

~\\
({\it c}) {\it Massless dilaton}: $\Lambda = 0$.

As was mentioned above recently this is the most studied case 
if $\Xi$ is a single exponent.
What about $\Xi$'s for our class?
We have from Eq. (\ref{e-cl1a}) the two cases
\ba
&&
\hat\Xi = 
a_1 e^{-\frac{\beta}{d_1}\phi}
-
\frac{2 k \beta / d_2}{\beta+ d_1^2} e^{d_1\phi}, \quad
\beta+ d_1^2 \not=0, \\
&&
\hat\Xi =2\, 
e^{i \sqrt\beta \phi}
\left(
a_1 +
\frac{i k \sqrt\beta}{d_2} \phi
\right), \quad
d_1 = i \sqrt\beta.
\ea
Note, if one assumes $d_1 = \pm\sqrt\beta$ then the first from
these equations might generate several 
trigonometric Maxwell-dilaton couplings if one 
adjusts $a_1$ ($d_2$), $Q$ and $P$.
Its solution for $d_1 = \sqrt\beta$ is 
($f\equiv \frac{a_1}{r} - \frac{k r}{d_2} $):
\be                                                      \lb{e-massl2}
e^U = c - d_2^2 f,
\ \
\frac{\omega - \omega_0}{4 d_2 Q P^2} = 
\int\! \frac{ \drm r / r}{f \pm \sqrt{f^2 - (4 P Q)^2}},
\ee
and $A, \phi$ are exactly as in Eq.(\ref{e-b2}).
Note that in single-charged cases the square root in $\omega$ 
disappears so the latter can be resolved in 
ordinary functions.

~\\
({\it d}) $\Lambda = a_2 \sinh^2 \phi$.

This potential is also of interest for string theory
though as a trial one.
Its qualitative study was done in Ref. \cite{gh93}
hence an exact solution is just in time.
First consider the singular subclass 
(\ref{e-cl1sb}).
Choosing $d_1 = i \sqrt\beta$ we have two subcases
(here $f\equiv 2 i \cosh \phi + \sqrt\beta \sinh \phi$):
\ba
&&
\hat\Xi = 
\frac{a_2  f^2
      e^{2 i \sqrt\beta \phi}
     }
     {d_2^2 (\beta +4)}
+
e^{i \sqrt\beta \phi}
\left(
a_1 +
\frac{2 i k \sqrt\beta}{d_2} \phi
\right)
, 
\\
&&
\hat\Xi = 
e^{-2\phi}
\biggl[
a_1 -
\frac{a_2+ 4 k d_2}{d_2^2} 
\phi
-
\frac{a_2}{2 d_2^2} 
e^{-2\phi}
\biggr]
, \ 
\beta = -4
.
\ea
Further, at arbitrary $d_1 \not= \pm i \sqrt\beta$ the 
equation (\ref{e-cl1a}) 
produces so ugly $\hat\Xi$  
that there is no sense to present it here.
Instead we tried to find some simple case 
more or less
resembling 
the string model of Ref. \cite{gh93} $\Xi \sim e^{-2\phi}$,
$\beta=4$.
For instance, if we assume the purely magnetic case and
impose $d_1=\sqrt\beta$, $\beta=4$ we obtain the model
which approximates the string one at large $P$ or $d_2$
\ba
&&
\Xi=
a_1 e^{-2\phi} +
\frac{a_2 e^{4\phi} (4 - 3 e^{2\phi})}{48 d_2^2 P^2}
-
\frac{k e^{2\phi}}{2 d_2 P^2}
,
\
e^{2\phi}=r,
\nn\\&&
e^U = c + k d_2 r + \frac{a_2 r}{24} (r^2 - 4 r +6)
- \frac{2  d_2^2 P^2 a_1}{r},
\ea
and $A$ is exactly as in Eq. (\ref{e-b2}).

~\\
({\it e}) {\it Quadratic potential}: $\Lambda = a_2 \phi^2$.

This classical potential 
(whose study in gravity can be traced back as far as Ref. \cite{fis48})
nowadays has been revived  as a test one
in string theory \cite{hh93,gh93}, 
but all the studies so far
were conducted at numerical or qualitative level only. 
For the sake of brevity
we consider only the case $\beta+ d_1^2 \not=0$ then
Eq. (\ref{e-cl1a}) yields
\ba
\hat\Xi = 
\frac{a_1}{e^{\frac{\beta}{d_1} \phi}}
+
  \frac{
        (\phi - \phi_+)
        (\phi - \phi_-)
       }
       {\alpha^3 d_2^2/ a_2}
e^{2 d_1 \phi}
-
\frac{2 k \beta/ d_2 }{\beta+d_1^2}  e^{d_1 \phi}
, 
\ea
where
$
\alpha \equiv \beta + 2 d_1^2 \not= 0$,
$\phi_\pm =  \frac{2d_1^2}{\alpha\beta}
(\pm \sqrt{\beta+d_1^2}-d_1 )$.
Again, the solution is unnecessary bulky so it is better
to assume $d_1=\sqrt\beta$, $\beta=1$ (we still have rescaling
freedom due to $a_2$), then we obtain:
\ba
e^U = c +
k d_2 r + 
\frac{a_2 r^2}{54} 
\bigl(
19+6 \ln r \, 
(3 \ln r -5)
\bigr)
-\frac{a_1 d_2^2}{r},   \nn
\ea
with $A$ and $\phi$ being given by Eq. (\ref{e-b2}),
and $\omega$ is  as in Eq. (\ref{e-massl2}) if
$
f \equiv \frac{a_1}{r} - \frac{k r}{d_2} -
\frac{a_2 }{27 d_2^2}r^2 (9 \ln^2 r + 12\ln r -4)
$.

Of course, we have mentioned just a few examples.
Other $\Xi$ or $\Lambda$ that might appear from a concrete
problem can be  paired up within our class in a similar manner.
Despite this pairing is artificial procedure the
generated exact 
solutions are better than numerical studies from
scratch, besides ones can verify or falsify qualitative
approaches and results.

{\it Moreover: new classes and full separability}.
The class discussed above   is not the only one  - 
other integrability classes of
the EMD system can be generated even without any use of
\cite{zlo-way} if we 
switch the independent variable from $r$ to $\phi$
(due to dilaton being an invertible function of $r$)
and 
relax the ansatz $A \sim \phi$ (\ref{e-cl1b}).
Suppose $A (r) \equiv \ap (\phi(r))$ 
then by linear rearrangement it
can be shown that the essential
system (\ref{e-em1}), (\ref{e-em3}) and (\ref{e-em4})
is equivalent to the following one 
\ba
&& 
2 k (p-1) + 
\frac{2 e^\ap}{p}
\left(
      \Lambda + e^{-p \ap} \hat\Xi
\right)
+
e^{\up + 2 Y}
\nn\\&&\quad
\times
\biggl(
      \frac{\beta}{p \ap_{,\phi}^{2}} 
-
\frac{\up_{,\phi}}{\ap_{,\phi}} 
- \frac{p-1}{2}
\biggr)
= 0, \\
&& 
2 k (p-1) + 
\frac{2 e^\ap}{p}
\left(
      \Lambda + \frac{p }{2\beta} \Lambda_{,\phi} \ap_{,\phi}
\right)
+
e^{\up + 2 Y}
\biggl(\frac{1}{\ap_{,\phi}}
\biggr)_{,\phi}
\nn\\&&\quad 
+
\frac{2 e^{-(p-1) \ap}}{p}
\left(
      \hat\Xi + \frac{p }{2\beta} \hat\Xi_{,\phi} \ap_{,\phi}
\right)
= 0,                                               \lb{e-g2}\\
&&
\phi' - \frac{e^{Y-\ap/2}}{ \ap_{,\phi}} =0,  \quad
Y (\phi) = - \frac{\beta}{p} 
\int \frac{\drm \phi}{ \ap_{,\phi}} + Y_0,       \lb{e-g3}
\ea
where $U (r) \equiv \up (\phi(r))$ and $p\equiv D-2 = 2$
(everything in this section is applicable for arbitrary $D$ provided
$P\equiv 0$ at $D \not= 4$).
Consider the first two  equations 
which are perfect both for obtaining $\ap$ at given
$\Lambda$ and $\Xi$ and for the inverse problem,
i.e., for obtaining the $\Lambda$-$\Xi$ pairs corresponding
to a concrete $\ap$.
Note, if $\ap \sim \phi$ $(= d_1 \phi - \ln d_2)$ 
then Eq. (\ref{e-g2}) immediately reduces to (\ref{e-cl1a}) 
otherwise one must proceed in the following way.
First step: customize $\ap$ being any function of $\phi$, e.g., 
$\ln{(a \phi^b)}$ or
$\ln{(a \sin\phi)}$.
Second, 
eliminate $\up$ from the first two  equations using
the rest ones to
obtain the class equation, 
\ba                                            
&&
\frac{H_{,\phi}}{\ap_{,\phi}}  
+
\left(
\frac{\beta}{p \ap_{,\phi}^2} + \frac{p-1}{2}
\right)
H
+
k (p-1) 
\nn\\&& \qquad \qquad\qquad\qquad\quad
+ 
\frac{e^\ap}{p}
\left(
      \Lambda + e^{-p \ap} \hat\Xi
\right)=0,                         \lb{e-g1a}
\ea
where
$
H \equiv
\frac{1}{p (1/\ap_{,\phi})_{,\phi}}
\biggl[
k p (p-1) + 
e^\ap
\left(
      \Lambda + \frac{p }{2\beta} \Lambda_{,\phi} \ap_{,\phi}
\right)
+
e^{-(p-1) \ap}
\left(
      \hat\Xi + \frac{p }{2\beta} \hat\Xi_{,\phi} \ap_{,\phi}
\right)
\biggr]
,
$
that does not contain $Y$ as well so
at fixed  $\ap$ 
it becomes a {\it linear} second-order ODE
with respect to $\Lambda$ and $\hat\Xi$.
Final step: 
choose either $\Lambda$ or $\Xi$ 
then the class
equation recovers the missing partner of the pair;
$U (\phi)$ is algebraically given by  (\ref{e-g2});
other equations readily yield $\phi (r)$ hence $U(r)$, $A(r)$,
$\omega (r)$.  

Someone doing the direct task (solving $A$ at given
$\Xi$ and $\Lambda$) 
appreciates the fact that the static
EMD system has been thus completely separated:
Eqs.
(\ref{e-em1}) - (\ref{e-em5}) are equivalent to
the mutually non-involved
ODEs (\ref{e-g1a}), (\ref{e-g2}), (\ref{e-g3}) and (\ref{e-em5}) 
yielding, respectively, $A$, $U$, $\phi$ and $\omega$.
Unfortunately, the most important first one is a non-linear
third-order ODE so the direct task is still hard to be accomplished
without supplementary symmetries or assumptions.

\def\AnP{Ann. Phys.}
\def\APP{Acta Phys. Polon.}
\def\CJP{Czech. J. Phys.}
\def\CMPh{Commun. Math. Phys.}
\def\CQG {Class. Quantum Grav.}
\def\EPL  {Europhys. Lett.}
\def\IJMP  {Int. J. Mod. Phys.}
\def\JMP{J. Math. Phys.}
\def\JPh{J. Phys.}
\def\FP{Fortschr. Phys.}
\def\GRG {Gen. Relativ. Gravit.}
\def\GC {Gravit. Cosmol.}
\def\LMPh {Lett. Math. Phys.}
\def\MPL  {Mod. Phys. Lett.}
\def\NPh  {Nucl. Phys.}
\def\PhE  {Phys.Essays}
\def\PhL  {Phys. Lett.}
\def\PhR  {Phys. Rev.}
\def\PhRL {Phys. Rev. Lett.}
\def\PhRp {Phys. Rept.}
\def\NCim {Nuovo Cimento}
\def\TMF {Teor. Mat. Fiz.}
\def\prp {report}
\def\Prp {Report}

\def\jn#1#2#3#4#5{{#1}{#2} {\bf #3}, {#4} {(#5)}} 

\def\boo#1#2#3#4#5{{\it #1} ({#2}, {#3}, {#4}){#5}}



\begin{references}
\bibitem{txt-uw}
Once we explicitly know $\Xi(\phi)$, $\Lambda (\phi)$,
$A(r)$ and $\phi (r)$,
the  function 
$U(r)$ can  be easily found from Eq. (\ref{e-em1}) when substituting 
them as functions of radius 
-
then this equation becomes a first-order
linear ODE with respect 
to $e^U$ and hence it is always resolved in quadratures;
$\omega$ can be found by direct integration (\ref{e-em5}).
\bibitem{txt-otherd}
Here we work in 4D but the case of arbitrary $D$, which
is especially interesting if $k=0$ ($p$-branes), can be done
similarly.
Running ahead, 
despite the D-dimensional
analogue of Eq. (\ref{e-cl1a})  does not undergo sufficient changes, 
the extended Lambda-Maxwell duality appears to be broken at $D \not= 4$.
It is curious that electric-magnetic duality is also
broken if $D \not= 4$.
\end{references}
\end{document}